\documentclass[aps,prl,twocolumn,showpacs,preprintnumbers,amsmath,amssymb,superscriptaddress]{revtex4}

\usepackage{epsfig}
\usepackage{dcolumn}
\usepackage{bm}
\usepackage{color} 

\begin{document}

\preprint{Submitted to {\it Physical Review Letters}}

\title{Global Gyrokinetic Simulation of Tokamak Edge Pedestal Instabilities}
\author{Weigang Wan}
\author{Scott E. Parker}
\author{Yang Chen}
\affiliation{Department of Physics, University of Colorado, Boulder, Colorado 80309, USA}
\author{Zheng Yan}
\affiliation{University of Wisconsin-Madison, Madison, Wisconsin 53706, USA}
\author{Richard J. Groebner}
\author{Philip B. Snyder}
\affiliation{General Atomics, Post Office Box 85068, San Diego, California 92186, USA}
\date{\today}

\begin{abstract}
Global electromagnetic gyrokinetic simulations show the existence of near threshold conditions for both a high-$n$ kinetic ballooning mode (KBM) and an intermediate-$n$ kinetic version of peeling-ballooning mode (KPBM) in the edge pedestal of two DIII-D H-mode discharges. When the magnetic shear is reduced in a narrow region of steep pressure gradient, the KPBM is significantly stabilized, while the KBM is weakly destabilized and hence becomes the most-unstable mode. Collisions decrease the KBM's critical $\beta$ and increase the growth rate.
\end{abstract}
\maketitle

Present-day tokamak fusion experiments achieve high performance with a narrow edge particle and energy transport barrier at the plasma boundary, called an ``edge pedestal." In the edge pedestal, the plasma density and temperature gradients build up and eventually drive electromagnetic instabilities at intermediate and high toroidal mode numbers ($n$). Understanding the underlying physics of pedestal instabilities from first principles is critical for predicting the performance of future large experiments such as ITER \cite{Aymar01}. The pedestal can be crudely characterized by the height and width of the pressure profile. \citet{Snyder09,Snyder11} developed a model that successfully predicts the experimentally observed height and width of the pedestal by combining the linear threshold of two electromagnetic plasma instabilities in the pedestal region: the intermediate-$n$ magnetohydrodynamic (MHD) ``peeling-ballooning'' mode (PBM) and the high-$n$ kinetic ballooning mode (KBM). While the PBM has been well parameterized using ideal MHD theory, it has never, until now, been verified using more realistic kinetic calculations. The KBM, on the other hand, requires a kinetic model.

The KBM threshold is known to be important in core turbulence simulations~\cite{Snyder01,Waltz10}. Transport levels in core calculations become very high when approaching the KBM threshold. The KBM threshold has not been clearly identified in previous gyrokinetic edge simulations~\cite{Told08,Chen08,Wan11aps,Wang12,Fulton11aps}. However, experimental evidence of KBMs exist in DIII-D H-mode (High confinement mode) and quiescent H-mode experiments~\cite{Yan11pop,Yan11prl}, and the observed profiles closely correspond to a simplified calculation of KBM criticality over a wide range of parameters~\cite{Groebner10,Snyder11}. Recently, {\small GS2} simulations of MAST~\cite{Dickinson11,Dickinson12} indicated an even parity mode in the steep gradient region of the pedestal of the spherical tokamak and identified the mode as a KBM. The main signature of a KBM -- an electromagnetic mode with a $\beta$ threshold, has still yet to be demonstrated, until this Letter.

We present a self-consistent picture of both the KBM and PBM in the pedestal, and describe the conditions when each of them dominates. We show that the H-mode pedestal, just prior to the onset of observed Edge Localized Mode (ELM) instabilities, is very near the KBM threshold in global gyrokinetic simulations. In addition to the high-$n$ KBM, an intermediate-$n$ electromagnetic mode is unstable and we identify it as a kinetic version of the MHD peeling-ballooning mode. Using the gyrokinetic $\delta\!f$ particle-in-cell code {\small GEM}~\cite{Chen03,Chen07} with electron-ion collisions, we study the global linear stability of H-mode pedestal profiles from two DIII-D experiments: discharge 136051 that has been previously reported~\cite{Yan11pop} with characteristics of KBM, and another discharge 132016. Calculations using these ``original'' profiles show two types of instabilities: an intermediate-$n$ mode that propagates in the electron diamagnetic direction in the plasma frame (we will call this mode the ``kinetic peeling ballooning mode'', KPBM) and a high-$n$, low frequency mode that mostly propagates in the ion direction (we refer to this mode as the ``ion mode''). These two modes are driven by the pressure gradient. While the ion mode has a ballooning structure, it lacks an important property of the KBM: the $\beta$ parameter scan should show a sudden change of real frequency corresponding to strong increase of growth rate (see, e.g. Refs~\cite{Falchetto03,Candy05}). Additionally, this ion mode is subdominant to the KPBM. The KPBM is very sensitive to the shape of the $q$ profile and can only be seen in global simulations and not in flux tube simulations. If we slightly manipulate the magnetic shear near the steep pressure gradient region of the pedestal, i.e., very locally flatten the $q$ profile, the KPBM would be significantly stabilized and the ion mode would be weakly destabilized and begin to show clear KBM characteristics. These results indicate that an improved pedestal model should include, in detail, any corrections to the bootstrap current~\cite{Wilson92,Sauter99,Kagan10}, and any other equilibrium effects that might reduce the local magnetic shear~\cite{Zhu12,Callen10}.

\begin{figure}[htp]
\epsfig{file=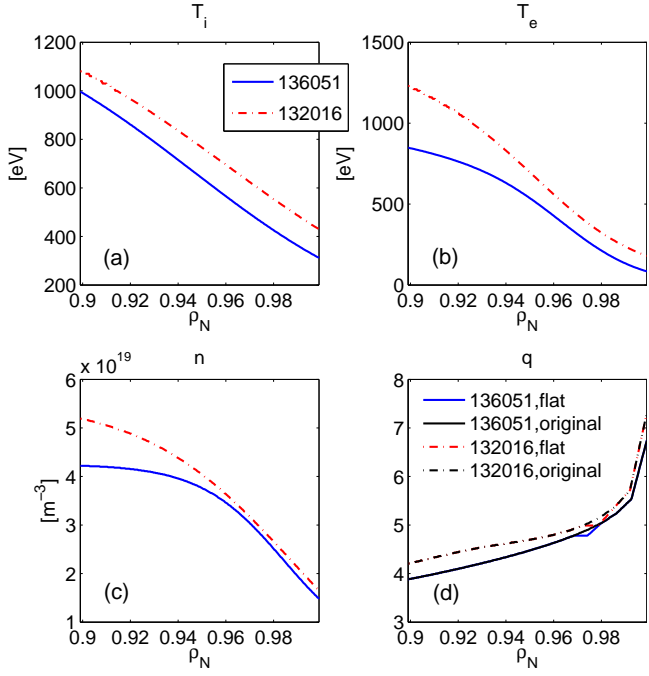,clip=true,width=8.5cm}
\caption{\label{eq}(Color online). The experimental profiles of (a) $T_i$, (b) $T_e$, (c) density $n$ and (d) original and flat safety factor $q$ for 136051 (solid lines) and 132016 (dash-dotted lines).}
\end{figure}

The two experimental profiles are shown in Fig.~\ref{eq}. The simulation box covers the $0.899\leq\rho_N\leq 0.999$ region inside the separatrix, where $\rho_N$ is the normalized radius. Fixed boundary conditions are applied and the density and temperature profiles are smoothed at the boundaries in simulation. The magnetic equilibria are parameterized using Miller equilibrium~\cite{Miller98,Chen08}. The magnetic equilibrium used for 136051 did not include corrections for the bootstrap current. The equilibrium for 132016 included corrections using the Sauter model~\cite{Sauter99}. The simulation domain grid is $64\times 32\times 32$, with $64$ cells along the radial direction. The time step is $\Delta t=1/\Omega_i$ where $\Omega_i$ is the proton gyrofrequency calculated at top of the pedestal. There are $1048576$ particles per species with realistic deuterium to electron mass ratio. Figure~\ref{eq}(d) shows the flattened $q$ profiles as well. 

\begin{figure}[htp]
\epsfig{file=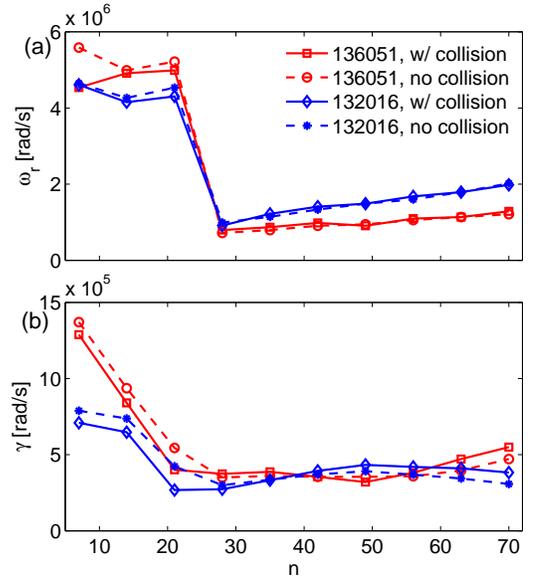,clip=true,width=6.8cm}
\caption{\label{orig}(Color online). The results of (a) linear real frequency $\omega_r$ and (b) growth rate $\gamma$ for the two discharges with original $q$ profiles. The effect of collisions is also shown.}
\end{figure}

\begin{figure}[htp]
\epsfig{file=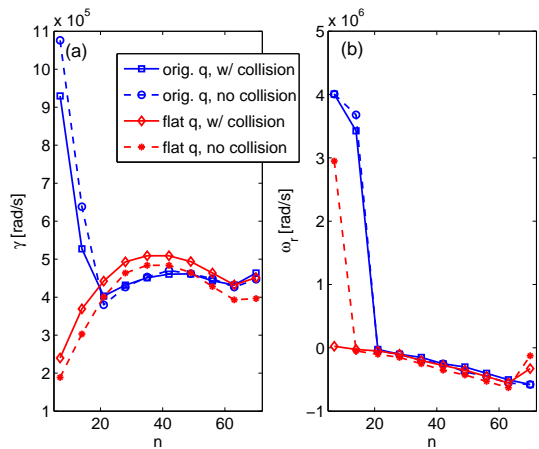,clip=true,width=7cm}
\caption{\label{gmomfo}(Color online). Simulations of 136051 at a high $\beta$ and without $E_r$. Results with the flat $q$ profile are compared to that with the original $q$, including the effect of collisionality.}
\end{figure}

\begin{figure}[htp]
\epsfig{file=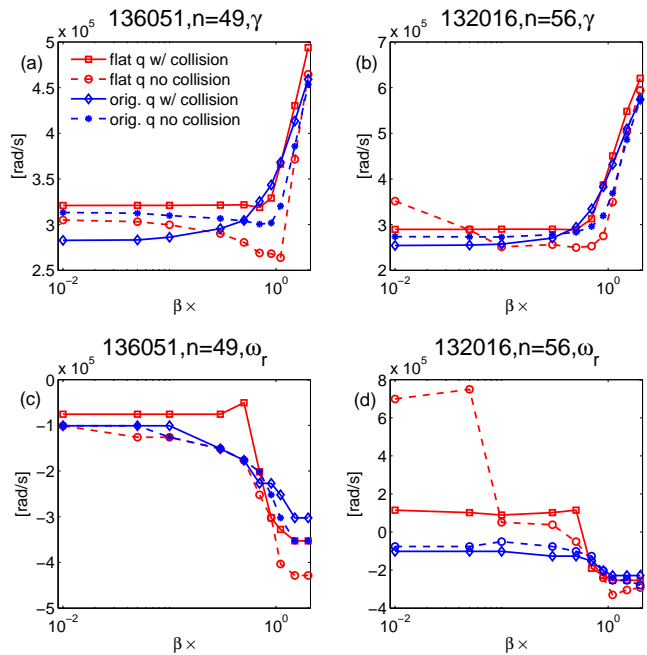,bb=23 116 571 675,clip=true,width=8.5cm}
\caption{\label{betax}(Color online). The $\beta$ scan of 136051 at $n=49$ and 132016 at $n=56$, with flat vs. original $q$ profiles. ``$\beta \times$'' means the factor that is used to multiply the experimental $\beta$.}
\end{figure}

Figure~\ref{orig} scans instabilities with mode number of $7\leq n\leq 70$ for the two profiles with the original $q$ profiles from experiments. The corresponding $k_y\rho_D$ at the center of the simulation box is in the range of $[0.102, 1.02]$ for 136051 and $[0.109, 1.09]$ for 132016, where $\rho_D$ is deuterium gyroradius. The two discharges exhibit quite similar trends. From Fig.~\ref{orig}(a) there are clearly two types of instabilities: intermediate-$n$ ($n\leq 21$) modes and high-$n$ modes. Both modes appear to propagate in the electron diamagnetic direction here indicated by their positive real frequencies. However, there is a Doppler shift caused by the radial electric field $E_r$ here. In simulations without $E_r$ shown later in Figs.~\ref{gmomfo} and~\ref{betax}, the high-$n$ instability propagates in the ion direction while the intermediate-$n$ instability still in the electron direction. The experiment of 136051~\cite{Yan11pop} has found two bands of density fluctuations, with an ion band at $50$ - $150$ kHz and an electron band at $200$ - $400$ kHz. Here for 136051, in the ``laboratory'' frame, the high-$n$ instability has a frequency equivalent to $160$ kHz, quite close to the frequency of the ion band found in experiment. However, unlike in the experiment, this mode is not the dominant instability here. As shown in Fig.~\ref{orig}(b), the intermediate-$n$ electron instability has a much higher growth rate. Its frequency is around $800$ kHz here, about twice that of the electron band of the experiment. 

\begin{figure}[htp]
\epsfig{file=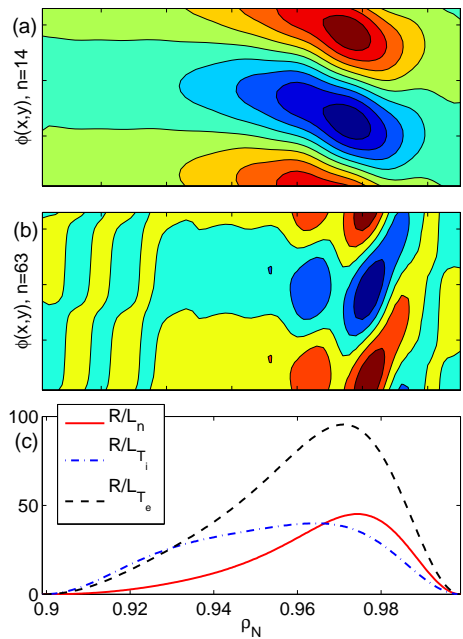,clip=true,width=6.cm}
\caption{\label{secgrad}(Color online). The mode structure (electrostatic potential contour plots) of the intermediate-$n$ (a) and high-$n$ (b) modes of 136051, with the original $q$ profile, and compared to the temperature and density gradients (c).}
\end{figure}
The electrostatic potential $\phi(x,y)$ of these two instabilities are shown in Fig.~\ref{secgrad}(a,b) for 136051, with $x$ and $y$ corresponding to the radial and toroidal direction in the field-line-following coordinate, respectively. Figure~\ref{secgrad}(c) also shows the temperature and density gradients, represented by $R/L_n$ and $R/L_T$, where $R$ is the major radius and $L_n^{-1}=d\ln n/dr$ and $L_T^{-1}=d\ln T/dr$. Note that the gradients are zero at the boundaries because the profiles are smoothed. The $x$ axes of the contour plots corresponds to the radius of Fig.~\ref{secgrad}(c). Both instabilities peak in the steep gradient region, indicating they are driven by pressure gradients. Both instabilities also have a largely even parity structure. However, the intermediate-$n$ mode appears to have a ``tail'' tilted towards the top of the pedestal; the dominant structure of the high-$n$ mode has a tail tilted towards the separatrix. In flux tube simulations we find ITG is the dominant instability on top of the pedestal, but its growth rate is weaker than that of the high-$n$ mode in the steep gradient region and therefore ITG is never dominant in global simulations. The possibility of trapped electron mode~\cite{Ryter05} is excluded because collisions don't decrease the linear growth rate. 

The results of the high-$n$ instability (ion mode) here agree with our flux tube simulations. In simulations with another profile 131997~\cite{Wan11aps}, a similar mode was found and the results agree with electrostatic simulations of {\small GTC}~\cite{Fulton11aps} and the flux tube eigenmode results of {\small GYRO}~\cite{Wang12}, although the mode has a positive real frequency for $k_y\rho_D<0.35$ without $E_r$ and was thus identified as an ``electron mode''. The intermediate-$n$ mode, which we now refer to as KPBM, is not observed in flux-tube simulations and is very sensitive to the $q$ profile. Measuring the pedestal $q$ profile is a difficult experimental challenge. Thus, experimental values for the q-profile are usually obtained from application of bootstrap current models to measured pedestal density and temperature profiles. In fact, the bootstrap current can vary from the generally used Sauter model~\cite{Kagan10}, and there is significant uncertainty in the measured gradients required to calculate the bootstrap current. In previous simulations of discharge 98889~\cite{Wan11pop}, which has a near-zero magnetic shear in a region across the steep gradient area~\cite{Callen10}, the KPBM is not present. We now flatten the $q$ profiles in a {\it very small} region in the two discharges as shown in Fig.~\ref{eq}(d). In doing so, we run flux tube simulations first to find a position in the steep gradient region that is locally most unstable, and then change the $q$ profile at that location with a zero magnetic shear.

\begin{figure}[htp]
\epsfig{file=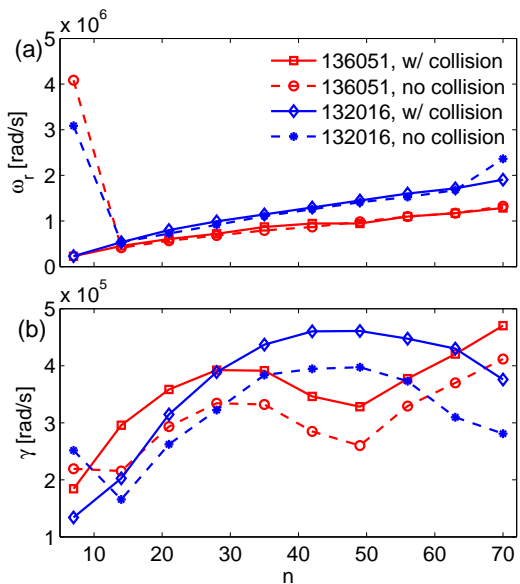,clip=true,width=6.8cm}
\caption{\label{flat}(Color online). Same as Fig.~\ref{orig} but with flattened $q$'s.}
\end{figure}

Figure~\ref{flat} shows the results with the ``flat'' $q$ profiles. The intermediate-$n$ KPBM is significantly stabilized and the high-$n$ ion mode, which we now identify as KBM, now dominates. The flat $q$ reduces the real frequency for $n=14$ and $n=21$, making the frequencies comparable with KBM. Collisionality further suppresses the high frequency of the $n=7$ mode and reduces its growth rate. In addition, the $\phi(x,y)$ mode structures of the modes (not shown) are also changed from Fig.~\ref{secgrad}, the tilted structure is reduced and the modes exhibit the more typical even structure. If we reduce the magnetic shear gradually instead of using a flat $q$ here, the growth rate of the KPBM is also reduced gradually.

We can see both the effects of the flat $q$ and collisionality more clearly in runs at higher values of $\beta$, where the modes are more unstable. Figure~\ref{gmomfo} shows results of 136051 with twice the experimental $\beta$ for both the original and flat $q$ profile. $E_r$ is removed to eliminate the Doppler shift. Although $E_r$ is generally believed to be a stabilizing factor at pedestal, here in linear simulations it is found to be destabilizing for the KPBM, thereby the growth rates in Fig.~\ref{gmomfo}(a) are smaller than in Fig.~\ref{orig}(b). The KBM now has a negative real frequency. It becomes obvious that the flat $q$ significantly stabilizes the KPBM and reduces its real frequency, leaving the KBM dominant. The KBM is moderately destabilized by the flat $q$, with its real frequency almost unchanged. Collisions reduce and even suppress the frequency of the KPBM, and are slightly destabilizing for the KBM. For $n>70$ the growth rate may rise again, but we restrict this study to $k_y\rho_i \leq 2$ where our gyrokinetic simulations are valid.
 
Figure~\ref{betax} shows a $\beta$ scan for the KBM for both experimental profiles. In simulations with both the slightly flattened $q$ and collisions, the instabilities of both profiles display the standard KBM features. As $\beta$ increases, the growth rate remains low (electrostatic), and then after passing the critical $\beta$, the growth rate strongly increases, with a corresponding sudden change in the real frequency with phase velocity in the ion direction. Since the experimental $\beta$ ($\beta \times=1$) is well above the critical $\beta$, the KBM is unstable, and is indeed the dominant instability, which in turn would limit the pressure gradients of the pedestal. The effect of collisions is to make the critical $\beta$ smaller. In simulations with the original $q$ profiles, there's not the characteristic sudden change in the real frequency, but the mode could still be identified as a modified KBM based on the linear results presented here. 

Resistive MHD studies of the PBM mode by~\citet{Zhu12} have shown that a flat $q$ stabilizes the PBM, the same property we see here with the KPBM. Besides being sensitive to the $q$ profile, the PBM is also electromagnetic, and in Ref.~\cite{Zhu12} it has $n\leq11$, a similar range of mode number of our KPBM. The unique property of the KPBM here is that it has a phase velocity in the electron diamagnetic direction while resistive MHD would show a near zero real frequency. 

Theoretically, the effects of magnetic shear (characterized by $\hat{s}=d\ln q(\rho)/d\ln\rho$) and pressure gradient (characterized by $\alpha=q^2R\beta/L_p$, where $L_p=p/\nabla p$ is the pressure scale length) on the stability of ballooning modes have been studied extensively. For core plasmas it is shown that the ideal MHD ballooning modes could be destabilized by a near-zero magnetic shear, as in the internal transport barrier~\cite{Conner04}; while KBMs are found to be unstable with a negative~\cite{Hirose96} or near-zero magnetic shear~\cite{Hirose03}. In the steep gradient region of the edge pedestal, however, the pressure gradient is so high and $\alpha$ is usually bigger than the stable threshold of typical core plasma $\hat{s}$ - $\alpha$ diagrams and the cases we studied here should be near the second stability region. The $\hat{s}$ - $\alpha$ analysis of edge plasmas could therefore be very different. Furthermore, our kinetic simulations suggest the gradients of density and electron and ion temperatures have different destabilizing effects, which apparently cannot be represented by a single HMD parameter $\alpha$. 

Transport codes~\cite{Kessel07}, MHD calculations~\cite{Zhu12} and experimental measurements~\cite{Groebner10} have all shown that the bootstrap current can flatten the $q$ profile. The kinetic linear stability of the edge pedestal is thus a subtle competition between the PBM and KBM as seen both here and previously in the EPED model~\cite{Snyder09,Snyder11}. Both are driven by the pressure gradients and therefore limit the pedestal shape. If the magnetic shear is high, PBM is much more unstable than KBM. Reducing the magnetic shear stabilizes PBM and KBM becomes the dominant instability. It is thus important to incorporate accurate
representations of the bootstrap current and edge geometry, as both strongly impact magnetic shear. Additionally, better experimental characterization of the edge $q$ profile would help test and improve predictive models.

This work is part of the Center for Plasma Edge Simulation supported by the Department of Energy Scientific Discovery through Advanced Computing program. Some work was supported by the U.S. Department of Energy under DE-FG02-89ER53296, DE-FG02-08ER54999, DE-FC02-04ER54698, and DE-FG02-95ER54309. We thank Choong-Seock Chang, James Callen, Eric Wang and Scott Kruger for useful discussions.


\end{document}